# DO YOU NEED A DAO?
# A FRAMEWORK FOR ASSESSING DAO SUITABILITY

*Short Paper*

Henrik Axelsen, University of Copenhagen, Copenhagen, Denmark, heax@di.ku.dk
Johannes Rude Jensen, University of Copenhagen, Copenhagen, Denmark, j.jensen@di.ku.dk
Omri Ross, University of Copenhagen, Copenhagen, Denmark, omri@di.ku.dk

## Abstract

*Decentralized Autonomous Organizations (DAOs) have seen exponential growth and interest due to their potential to redefine organizational structure and governance. Despite this, there is a discrepancy between the ideals of autonomy and decentralization and the actual experiences of DAO stakeholders. The Information Systems (IS) literature has yet to fully explore whether DAOs are the optimal organizational choice. Addressing this gap, our research asks, "Is a DAO suitable for your organizational needs?" We derive a gated decision-making framework through a thematic review of the academic and grey literature on DAOs. Through five scenarios, the framework critically emphasizes the gaps between DAOs' theoretical capabilities and practical challenges. Our findings contribute to the IS discourse on blockchain technologies, with some ancillary contributions to the IS literature on organizational management and practitioner literature.*

## 1    Introduction

Decentralized Autonomous Organizations (DAOs) are organizations run by rules encoded as computer programs called smart contracts. They have gained traction in recent years as a preferred organization for token-based communities operating on blockchain. Data suggests that more than 12,000 DAOs (Rikken et al. 2023) control around US$19Bn in their treasuries, owned by some 9m token holders, of which 3m are active voters (DeepDAO 2023). Despite the continuing regulatory uncertainty and the persistence of debilitating hacks and exploits, the number of active DAOs appears to grow by double-digits year after year (Bellavitis et al. 2023). This suggests practitioners continue finding compelling use cases for this novel style of organizing.

Yet, recent empirical data indicate a considerable gap between the espoused values and practical realities of DAOs (Feichtinger et al. 2023). Given the clear profit motive in issuing so-called governance tokens, critics argue that the DAO moniker is too widely used by organizations that are neither decentralized nor autonomous in practice.

While there is a growing body of multidisciplinary literature on DAOs within IS, little work has been done to ascertain whether a DAO is actually the right organizational model for governing a given product or service. This research-in-progress (RiP) paper addresses this gap in the Information Systems (IS) literature on blockchain technology. We ask the research question: "Is a DAO suitable for your organizational needs?" The research question is formulated as a binary choice, resulting in a 5-step gated decision-making framework derived from a thematic analysis of the academic and grey literature on DAO definitions and capabilities. While our contribution to the IS discourse on blockchain technologies and DAO is theoretical, the work presented in this RiP paper may also carry relevance to the IS literature on organizations (Augustin et al. 2023; Pohl et al. 2022).

## 2      Method

To understand how the literature developed in the short time span in which DAOs have been around as an organizational category, we conducted a scoping review of the literature (Arksey and O'Malley 2005) followed by a thematic analysis of definitions and organizational implications (Xiao and Watson 2019) through a socio-technical lens. Initiated by our research question, we first targeted search for literature across AIS eLibrary, IEEE, ACM electronic library, SCOPUS, Elsevier's Science Direct and Springer on "Decentralized Autonomous Organization." This revealed more than 19,000 results on the first five databases and more than 44,000 on Springer. Across all databases, most results were unrelated to DAOs. We then added "DAO" to the search string, which reduced the search to 1,831 and 2,546 results, respectively. Removing duplicate entries narrowed the result to a combined 3,102 results. These were sorted by relevance (weighing the full text of each document, where it was published, who it was written by, and how often and how recently it has been cited in other scholarly literature) and reviewed manually. We excluded technical papers around non-DAO autonomous systems, purely descriptive papers, and papers focusing on commercial DAO platforms or specific applications of DAOs. We then prioritized manuscripts that explored DAO definitions, typologies, and their relevance to organizational contexts, ending the search with 18 results. We then conducted a similar search on Google Scholar and cross-checked the 18 results backward and forward. Sorting by relevance, this led to a further 20 results.

As the development of concepts may include elements of practical insights (Gregor et al. 2020), we then complemented the review with 'pragmatic inference' through a grey literature search on the same search string to capture additional insights from online networks. As grey literature is not subjected to peer review, we devised a protocol for systematically evaluating the material and determining inclusion in the literature review. We lean on (Gramlich et al. 2023) in ascertaining the credibility of the grey literature. This final search added an additional three results, totaling 41 results. Aiming to provide a balanced view that encompasses emerging academic and practical dimensions, the search culminated in 30 papers with unique contributions to DAO definitions, which we included in Table 1, organized by theme.

## 3      Thematic Analysis of the Literature

This section features a thematic analysis of the academic and grey literature (Table 1). By synthesizing the literature, we arrive at nine themes. In recent years, the literature on DAOs has become increasingly multidisciplinary (Santana and Albareda 2022), reflecting the application of the concept across a range of sectors. Nevertheless, the primary application is associated with decentralized finance (DeFi) (Schueffel 2021). Within the IS literature, scholars consider DAOs the combination of (i) decentralized applications implemented as 'smart contracts' deployed on a public blockchain and (ii) a set of organizational bylaws directing human efforts associated with the DAO (Pohl et al. 2022; Wang et al. 2019). Implementations differ by the extent to which the organizational logic is automated *'on-chain'* by smart contracts or *'off-chain'* through legal or social agreements between stakeholders. This combination of social and technological elements has led to an array of new organizational types (Hsieh et al. 2018; Lumineau et al. 2021; Murray et al. 2021). A key point of contention in the literature is the degree to which control of the DAO can become captured by single stakeholders, challenging the ideal for 'decentralization' (Kitzler et al. 2023; Orrick 2023), which has already led some practitioners to abandon the DAO concept altogether (McConaghy 2022).

| Theme | Description |
|---|---|
| **Decentralization and distribution** | Decentralization, key in DAO definitions (Santana and Albareda 2022), highlights the absence of a central authority, with decision-making distributed spread among members (Hsieh et al. 2018) and a key differentiator to traditional firms (Buterin 2022). This concept of a trustless, permissionless structure has persisted, recently suggesting a distinction between decentralization (Vergne 2020) and distribution (Berg et al. 2019) with likely at least 20 token-holders for long-term survivability (Rikken et al. 2023). |

| **Autonomy and automation** | Autonomy is a key attribute starting from (Vitalik Buterin 2014) through (El Faqir et al. 2020; Wang et al. 2019). The concept of automation (van Rijmenam 2019) has become a prominent feature, as automation of business processes is considered key in reducing bureaucracy(Qin et al. 2023). |
|---|---|
| **Organization and operations** | DAOs are generally internet-native organizations, more recently meta-organizations (Mini et al. 2021), coordinated, owned, and managed by members (Bellavitis et al. 2023; Hassan and De Filippi 2021). Organizational design theory shows an increased understanding of the characteristics of DAOs as entities that coordinate collective action and decision-making (Pohl et al. 2022). |
| **Smart contracts and permissionless blockchains** | From (Wright and De Filippi 2015) to (Rozas et al. 2021), smart contract infrastructure deployed in permissionless blockchains has been recognized broadly as the only way to run a DAO. This continues today, where public and permissionless blockchains remain the favored execution environment for DAO stakeholders. Smart contracts implement the DAO decision-making tools and the associated governance token. |
| **Self-governance and code-based governance** | The concept of 'self-governance' through code, starting from (Jentzsch 2016), is a recurrent theme from (Davidson et al. 2018) to (Wright 2021) and (Ziegler et al. 2022) emphasizing the necessity of active stakeholder participation in both 'on-chain' and 'off-chain' activities (Santana and Albareda 2022). In recent literature (Zargham and Nabben 2022), the idea that rules are formalized, automated, and enforced by software appears consistently as a qualifier for DAOs. Today, most DAOs are governed through the use of 'governance-tokens,' small scripts enabling stakeholders to vote proposals in binary decision-making processes. |
| **Token economy and incentives** | Several definitions after 2016, like those of (Voshmgir 2017), emphasize token systems and economic incentives that coordinate distributed and fluid work practices within DAOs (Schirrmacher et al. 2021). Token-based incentives can be issued in the native governance token for the DAO, equivalent to employee stock options, or in stablecoins or other crypto-assets held in the DAO treasury. |
| **Human involvement** | Early definitions, like (Vitalik Buterin 2014), emphasize reliance on individuals, whereas later definitions (Filippi and Wright 2018) move towards a vision of complete automation by smart contracts and algorithms (Faqir-Rhazoui et al. 2021). The concept of full automation has become more pronounced in later years (Qin et al. 2023), while some still remain skeptical (El Faqir et al. 2020). |
| **The legal and formal structure** | While the very early definitions did not delve into the legal implications of DAOs, this topic has become increasingly relevant, starting with (Jentzsch 2016) and (Wright and De Filippi 2015) exploring legal personality and responsibility in taxonomy development to (Wright 2021) discussing participation without legal boundary. |
| **Scope and Potential** | As the literature matured, scholarship increasingly turned towards exploring the ultimate scope of DAO governance (Atzori 2017; DuPont 2019), some seeking to stretch the limits of what the concept may accomplish (Singh and Kim 2019) at which DAOs can operate, from small companies to meta-organizations (Wiriyachaokit et al. 2022). Recent definitions (Pahuja and Taani 2022; Qin et al. 2023) view DAOs as key to a so-called 'decentralized society.' |

*Table 1.      Key DAO themes.*

# 4     Do You Need a DAO?

Since the initial attempts at decentralizing authority using smart contracts, DAOs have evolved into complex entities, some featuring a legal and physical presence in multiple jurisdictions. This reflects a maturation from the original concept into what are now complex, multi-dimensional organizations.

Derived from the thematic literature review, we propose a forward-looking definition that aims to capture the theoretical and practical essence of DAOs:

> A DAO is a collaborative, open, blockchain-enabled platform governed by smart contracts designed to operate without centralized control. A DAO orchestrates interactions, asset management, and decision-making through coded rules to achieve common objectives, with global reach and integration with digital and virtual environments.

With this definition and the nine themes above, we propose a 5-step gated decision-making framework guided by the research question (Figure 1). The framework is designed to guide a decision-making process toward ascertaining whether or not a DAO is an appropriate choice for managing a service or product.

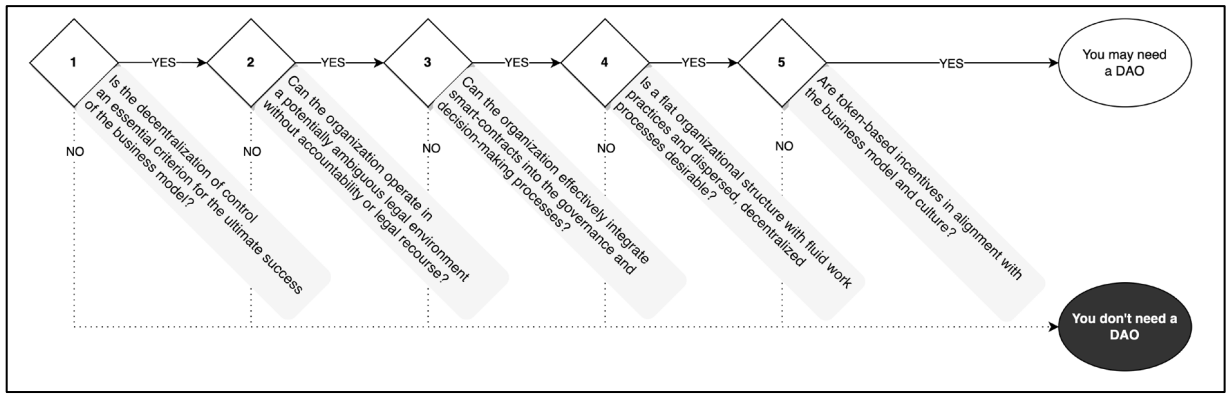

*Figure 1.    The 5-step Framework: Do you need a DAO.*

In this section, we break down the decision-making framework,

**Step 1: Is the decentralization of control an essential criterion for the ultimate success of the business model?** While DAOs do not technically require a specific type of blockchain, the consensus in the literature is on permissionless public blockchains. For a DAO to inherit the decentralized and permissionless properties of the underlying blockchain, it should have an active set of at least ~20 members. This step is vital to ensure diverse opinions and prevent risks associated with centralized control, like reduced transparency and manipulation (Rikken et al. 2023). To evaluate if a DAO suits your needs, assess how well the permissionless blockchain's capabilities and your organizational goals align with the concept of decentralization. This step involves determining if the technology aligns with the organization's mission (Pedersen et al. 2019) while also fulfilling requirements for decentralization – both technologically, through the blockchain's capabilities – and organizationally regarding decision-making and governance structures (Axelsen et al. 2022). A DAO is likely not needed if these conditions cannot be met.

**Step 2: Can the organization operate in a potentially ambiguous legal environment without accountability or legal recourse?** To evaluate the operational feasibility of a DAO within an ambiguous legal climate, it is essential to consider the evolving legal implications and regulatory uncertainties highlighted in the literature and recent judicial practice (Jentzsch 2016; Orrick 2023; Wright and De Filippi 2015). Understanding the legal context in which DAOs operate, including the potential for members' liability and the importance of legal personality and responsibility (Wright 2021), is critical. Stakeholders must understand the legal context in which the DAO will operate, particularly the implications for liability and regulatory compliance. This ensures that the DAO is prepared to navigate the legal complexities of operating on a blockchain, especially in a rapidly evolving regulatory landscape. Without incorporation and with no one to hold accountable, there is a risk that a DAO becomes treated as an unincorporated partnership, where each member of the DAO becomes personally liable for their actions, including voting. Aspiring DAO stakeholders must understand the legal context of the services offered and organize accordingly.

**Step 3: Can the organization effectively integrate smart contracts into the governance and decision-making processes?** The degree of autonomy and automation determines how much a DAO can operate independently from human coordination or control by embedding business logic in smart

contracts for governance and decision-making. The advantages of doing so include the assurance of deterministic execution and increased transparency, which may reduce bias in decision-making processes. Still, the implications of embedding governance and operational procedures within immutable code require thorough consideration, given the potential for accidentally introducing adverse incentives or excessive bureaucracy. The degree of automation in a DAO can vary depending on its objective function. For DAOs with a narrow and specific objective function, full automation may be an option, eliminating the need for organizational bylaws altogether (McConaghy 2022). In contrast, DAOs with broader objectives may benefit from a hybrid model incorporating automated processes and human involvement. Viewing DAOs as complex adaptive systems, stakeholders must balance scalability and stability when adapting to dynamic environments. While smart contract design can accommodate upgradeability and failsafe systems to protect against external and internal challenges, stakeholders should only proceed if the organization can benefit from the advantages of smart-contract-based organizational infrastructure.

**Step 4: Is a flat organizational structure with fluid work practices and dispersed, decentralized processes desirable?** Consider if a flat organizational structure with fluid work practice suits the organizational need. DAOs represent a departure from conventional hierarchical organizations, typically having very fluid entry/exit barriers and role designations (Schirrmacher et al. 2021). A flat organizational structure promotes direct, decentralized decision-making and collaboration in which teams and stakeholders are distributed across different locations, typically working asynchronously and remotely. Adopting a DAO model requires shifting from traditional hierarchies to a distributed operational and governance approach, entailing technological and cultural changes toward autonomy and remote collaboration.

**Step 5: Are token-based incentives in alignment with the business model and culture?** Various definitions post-2016, such as (Voshmgir 2017), emphasize the role of token systems and economic incentives in coordinating distributed work practices within DAOs. This aspect is critical for understanding how token-based incentives can align with or diverge from an organization's business model and culture. The literature presents a spectrum of perspectives on the role of human participants within DAOs, spanning reliance on individuals for certain tasks (Vitalik Buterin 2014) to later definitions, proposing complete automation (El Faqir et al. 2020). Stakeholders considering DAO governance must consider how token-based incentives might impact engagement and participation (Schirrmacher et al. 2021). Incentives drive behavior, which requires clear and quantifiable objectives. If these are not present, achieving anything may be difficult and frustrating compared to traditional organizations. Thus, careful consideration must be given to (i) whether incentive-driven work practices can be implemented without introducing apathy, (ii) the degree to which pseudonymity is an engagement problem, (iii) which voting-based decisions should be made immediately vs. longer term, and whether they are permanent or temporary; and (iv) how contributions should be assessed and rewarded; DAOs focusing on building a relationship-driven culture may reward mainly activities, others with a more transactional culture may primarily reward results.

# 5      Discussion

Through a thematic analysis of the academic and grey literature on DAOs, we propose a forward-looking definition of DAOs and derive a 5-step gated framework, guiding the decision-making on whether a DAO is the right choice as an organization. Our research contributes to the theoretical understanding of how DAOs integrate with organizational theory. It also serves as a practical guide, reflecting DAOs' potential as emergent organizational forms within the convergence of social and technological dynamics.

The concept of DAOs took root with the advent of Ethereum and improved smart contracts from 2014 onwards. That period marked the beginning of DAOs' practical applications (Jentzsch 2016), showcasing their potential and identifying some of their challenges (Wang et al. 2019). Since then, DAOs have evolved and now cover a range of applications across industries, but the challenges remain.

Our preliminary analysis indicates that DAOs may excel in scenarios where distributed governance and collective decision-making can harness diverse perspectives, democratizing the organizational process and potentially leading to more equitable and innovative outcomes. For industries or projects centered around collaborative open-source ventures, the transparency, audibility, and shared ownership facilitated by DAOs might enhance trust and alignment among stakeholders while overcoming their shortcomings in complexity. This may be particularly relevant where global participation is too difficult or costly through traditional organizational forms.

Our preliminary analysis also suggests multiple cases in which stakeholders in open-source and non-hierarchical organizations will <u>not benefit</u> from establishing a DAO: (1) Primarily, situations in which stakeholders have a centralized control preference or require centralized leadership and decision-making. This may be the case when managing stakeholders face regulatory and legal constraints, as the pseudonymous and permissionless nature of DAOs can conflict with industries or regions that demand clear accountability and specific legal structures. (2) Business models needing nuanced, expert-driven decision-making may find DAOs' reliance on smart contracts and member voting inadequate. If rapid adaptation to external shocks is needed, reliance on decentralized consensus and smart contracts may be too cumbersome. On the other hand, for businesses that do not require agility and rapid decision-making, the risk of member apathy or low engagement can become an issue when faulty implementation of incentives causes excessive bureaucracy or stakeholder apathy. (3) For business models with very narrow objective functions, automatable objectives may render the complexity of DAOs unnecessary. A DAO operating in a simple operational context could introduce security vulnerabilities or attract unneeded regulatory attention to an otherwise functional product. Thus, implementing a DAO for narrow objectives might lead to inefficiency and wasted resources. Even worse, in cases where operations can be automated, implementing a DAO could ironically lead to centralization, as the control might effectively rest with a small group of individuals who design and maintain the automation processes.

With the concept's gradual maturity, DAOs now represent diverse organizational solutions, leaving stakeholders to decide on the trade-offs between opposing organizational objectives. While DAOs may not fit traditional organizational models or policy objectives, their potential to redefine collaboration, ownership, and decision-making in the digital age offers a compelling solution within the identified limitations. This perhaps indicates a pathway toward more dynamic, inclusive, and resilient organizational forms.

Looking forward, DAOs may be poised to evolve towards more sophisticated governance models, focusing on ethical considerations and potential synergies with emerging technologies such as AI and IoT. Considering the challenges identified, the IS community must remain vigilant of technological advancements and regulatory shifts, underscoring our research's role in equipping practitioners with the knowledge to navigate these developments.

# 6      Conclusion

In this research-in-progress paper, we explored the DAO governance domain through a thematic analysis and literature review. We derive a forward-looking definition and a 5-step decision-making framework. Motivated by our research question, we then explore the challenges DAO governance may pose in a rapidly evolving digital and regulatory landscape.

Our findings offer actionable insights for navigating the increasingly complex technical landscape emerging in academic and practitioner literature. These insights are valuable for IS scholars and practitioners considering venturing into the field of DAOs because they can aid in informed decision-making regarding the feasibility and challenges of DAO governance. To this end, we contribute to the growing discourse on DAO within the IS literature on the application of blockchain technologies and organizations.

*Do You Need a DAO?*
in Decentralized Autonomous Organizations," *Proceedings of the 42nd International Conference on Information Systems*, December 12-15.

Murray, A., Kuban, S., and Anderson, M. J. J. 2021. "Contracting in the Smart Era: The Implications of Blockchain and DAOs for Contracting and Corporate Governance," *Academy of Management Perspectives* (35:4), pp. 622–641.

Orrick, W. H. 2023. *Case 3:22-Cv-05416-WHO CFTC vs Ooki DAO*, pp. 1–16.

Pahuja, A., and Taani, I. 2022. "From Constitution to Disbandment: Ephemeral Decentralized Autonomous Organizations Autonomous Organizations," *ICIS 2022 Proceedings*, pp. 0–9.

Pedersen, A. B., Risius, M., and Beck, R. 2019. "A Ten-Step Decision Path to Determine When to Use Blockchain Technologies," *MIS Quarterly Executive* (18:2), pp. 99–115.

Pohl, M., Degenkolbe, R., Staegemann, D., and Turowski, K. 2022. "Decentralised Autonomous Organisations in Organisational Design Theory," *MCIS 2022 Proceedings*.

Qin, R., Ding, W., Li, J., Guan, S., Wang, G., Ren, Y., and Qu, Z. 2023. "Web3-Based Decentralized Autonomous Organizations and Operations: Architectures, Models, and Mechanisms," *IEEE Transactions on Systems, Man, and Cybernetics: Systems* (53:4), IEEE, pp. 2073–2082.

van Rijmenam, M. 2019. *Sociomateriality in the Age of Emerging Information Technologies: How Big Data Analytics, Blockchain and Artificial Intelligence Affect Organisations*, (February).

Rikken, O., Janssen, M., and Kwee, Z. 2023. "The Ins and Outs of Decentralized Autonomous Organizations (DAOs) Unraveling the Definitions, Characteristics, and Emerging Developments of DAOs," *Blockchain: Research and Applications*, Zhejiang University Press, p. 100143.

Rozas, D., Tenorio-Fornés, A., Díaz-Molina, S., and Hassan, S. 2021. "When Ostrom Meets Blockchain: Exploring the Potentials of Blockchain for Commons Governance," *SAGE Open* (11:1).

Santana, C., and Albareda, L. 2022. "Blockchain and the Emergence of Decentralized Autonomous Organizations (DAOs): An Integrative Model and Research Agenda," *Technological Forecasting and Social Change* (182:June), Elsevier Inc.

Schirrmacher, N.-B., Jensen, J. R., and Avital, M. 2021. "Token-Centric Work Practices in Fluid Organizations: The Cases of Yearn and MakerDAO," *The 42nd International Conference on Information Systems* (December). (https://aisel.aisnet.org/icis2021).

Schueffel, P. 2021. "DeFi: Decentralized Finance - An Introduction and Overview," *Journal of Innovation Management* (9:3), I–XI.

Singh, M., and Kim, S. 2019. "Blockchain Technology for Decentralized Autonomous Organizations," *Advances in Computers* (115), Elsevier, pp. 115–140.

Vergne, J. 2020. "Decentralized vs. Distributed Organization: Blockchain, Machine Learning and the Future of the Digital Platform," *Organization Theory* (1:4), p. 263178772097705.

Vitalik Buterin. 2014. *DAOs, DACs, DAs and More: An Incomplete Terminology Guide*. (https://blog.ethereum.org/).

Voshmgir, S. 2017. "Disrupting Governance with Blockchains and Smart Contracts," *Strategic Change* (26:5), pp. 499–509.

Wang, S., Ding, W., Li, J., Yuan, Y., Ouyang, L., and Wang, F. Y. 2019. "Decentralized Autonomous Organizations: Concept, Model, and Applications," *IEEE Transactions on Computational Social Systems* (6:5), IEEE, pp. 870–878.

Wiriyachaokit, W., Augustin, N., Eckhardt, A., and Eckhardt, A. 2022. "Exploring Drivers of Sustained Participation in Decentralized Autonomous Organizations," in *ICIS 2022 Proceedings. 11*.

Wright, A. 2021. *The Rise of Decentralized Autonomous Organizations: Opportunities and Challenges*, (85:5), pp. 4–23. (https://letstalkbitcoin.com/is-bitcoin-overpaying-for-false-security.).

Wright, A., and De Filippi, P. 2015. "Decentralized Blockchain Technology and the Rise of Lex Cryptographia," *SSRN Electronic Journal*.

Xiao, Y., and Watson, M. 2019. "Guidance on Conducting a Systematic Literature Review," *Journal of Planning Education and Research* (39:1), pp. 93–112.

Zargham, M., and Nabben, K. 2022. *Aligning 'Decentralized Autonomous Organization' to Precedents in Cybernetics*.

Ziegler, C., Welpe, I. M., Taxonomy, A., and Welpe, I. 2022. *A Taxonomy of Decentralized Autonomous Organizations*, pp. 0–17. (https://aisel.aisnet.org/icis2022).
*Thirty-Second European Conference on Information Systems (ECIS 2024), Paphos, Cyprus*     8